\RequirePackage{lineno}
\documentclass{nature}
\usepackage{xcolor}
\usepackage{amsmath}
\usepackage{caption}
\usepackage{graphicx}
\newcommand{\unit}[1]{\,\mathrm{#1}} 
\newcommand{\equa}[1]{Eq.~\eqref{#1}}

\makeatletter
\let\saved@includegraphics\includegraphics
\AtBeginDocument{\let\includegraphics\saved@includegraphics}
\renewenvironment*{figure}{\@float{figure}}{\end@float}
\makeatother

\title{Floquet engineering enabled by charge density wave transition}

\author{Fei Wang$^{1,2,\ast}$, Xuanxi Cai$^{1,2,\ast}$, Teng Xiao$^{1,2,\ast}$, Changhua Bao$^{1,2}$, Haoyuan Zhong$^{1,2}$, Wanying Chen$^{1,2}$, Tianyun Lin$^{1,2}$, Tianshuang Sheng$^{1,2}$, Xiao Tang$^{1,2}$, Hongyun Zhang$^{1,2,3}$, Pu Yu$^{1,2}$, Zhiyuan Sun$^{1,2,\dagger}$ \& Shuyun Zhou$^{1,2,\dagger}$}

\usepackage{graphicx}
\makeatletter
\let\saved@includegraphics\includegraphics
\AtBeginDocument{\let\includegraphics\saved@includegraphics}

\makeatother

\begin{document}
\maketitle

\begin{affiliations}

\item Department of Physics, Tsinghua University, Beijing 100084, People's Republic of China
\item State Key Laboratory of Low-Dimensional Quantum Physics, Tsinghua University, Beijing 100084, People's Republic of China
\item Advanced Institute for Materials Research (WPI-AIMR), Tohoku University, Sendai 980-8577, Japan

* These authors contributed equally to this work\\
$\dagger$ Correspondence and request for materials should be sent to zysun@tsinghua.edu.cn, syzhou@mail.tsinghua.edu.cn

\end{affiliations}



\begin{abstract}

Floquet engineering has emerged as a powerful approach for dynamically tailoring the electronic structures of quantum materials\cite{oka2019floquet,rudner2020NRP,ZhouNRP2021,Sentef2021} through time-periodic light fields generated by ultrafast laser pulses. The light fields can transiently dress Bloch electrons, creating novel electronic states  inaccessible in equilibrium\cite{oka2009PRB,demler2011PRB,lindner2011NP,
Podolsky2013PRL,claassen2016NC,yan2016PRL,zhang2016PRB,Rubio2017NC}. 
While such temporal modulation provides dynamic control, spatially periodic modulations, such as those arising from charge density wave (CDW) order, can also dramatically reconstruct the band structure through real-space symmetry breaking. The interplay between these two distinct forms of modulation—temporal and spatial—opens a new frontier in electronic-phase-dependent Floquet engineering. 
Here we demonstrate this concept experimentally in the prototypical CDW material 1T-TiSe$_2$. Using time- and angle-resolved photoemission spectroscopy (TrARPES) with mid-infrared pumping, we observe a striking pump-induced instantaneous downshift of the valence band maximum (VBM), which is in sharp contrast to the subsequent upward shift on picosecond timescale associated with CDW melting. Most remarkably, the light-induced VBM downshift is observed exclusively in the CDW phase and only when the pump pulse is present, reaching maximum when pumping near resonance with the CDW gap. These observations unequivocally reveal the critical role of CDW in the Floquet engineering of TiSe$_2$.
Our work demonstrates how time-periodic drives can synergistically couple to spatially periodic modulations to create non-equilibrium electronic states, establishing a new paradigm for Floquet engineering enabled by spontaneous symmetry breaking.

\end{abstract}

\renewcommand{\thefigure}{\textbf{Fig.~\arabic{figure}}}
\setcounter{figure}{0}


Floquet engineering, the coherent manipulation of quantum states via time-periodic driving, offers a powerful approach to tailor electronic structures of quantum materials and induce novel quantum phases of matter\cite{oka2019floquet,rudner2020NRP,ZhouNRP2021,Sentef2021}. Over the past decade, a wide range of light-induced phenomena have been predicted, such as light-induced quantum anomalous Hall effect in graphene\cite{oka2009PRB,demler2011PRB}, Floquet topological insulators in semiconductors\cite{lindner2011NP,Podolsky2013PRL,claassen2016NC}, and light-induced phase transitions in topological semimetals\cite{yan2016PRL,zhang2016PRB,Rubio2017NC}. Experimentally, Floquet engineering leads to dressing of the electronic structures forming Floquet-Bloch states\cite{Gedik2013,Gedik2016,zhou2023pseudospin,HuberNature2023,Mahmood2025floquet,stfan2025graphene,gedik2025graphene}, as well as modifications of optical and transport properties\cite{GedikOpticalStark2015,CavalleriGrapheneFloquet2020,Hsieh2021nat,lee2022nat,Kogar2024Cr2O3}. Light-induced renormalization of the transient electronic structures has been reported in selected systems, such as the surface states of topological insulators\cite{Gedik2013,Gedik2016,HuberNature2023,Mahmood2025floquet} and semiconducting black phosphorus\cite{zhou2023pseudospin,ZhouBPPRL2023}, providing direct insights into the dynamic control of quantum materials.

A new frontier in non-equilibrium control is achieving Floquet engineering in systems with spontaneous spatial periodicity, exemplified by charge density wave (CDW) phases. These states develop spatially periodic lattice distortions that break the translational symmetry and profoundly modify the electronic structure. The interaction between such static spatial order and time-periodic light fields creates a novel regime of electronic-phase-dependent Floquet engineering, where light-driven band modifications become intrinsically linked to the symmetry-broken state. 

Here, we demonstrate Floquet engineering in 1T-TiSe$_2$, which is intrinsically enabled by the spatially modulated CDW phase, using time- and angle-resolved photoemission spectroscopy (TrARPES) with mid-infrared (MIR) pumping. Unlike far-above-gap pumping at 800 nm which induces an upward shift of the spectral peaks consistent with CDW melting, near-resonance pumping with the CDW gap leads to an instantaneous downshift of the valence band maximum (VBM). The light-induced band renormalization occurs concurrently with the pump pulse, decreases with increasing sample temperature and vanishes above the CDW transition temperature $T_{\text{CDW}}$ or when the pump photon energy is detuned from resonance with the CDW gap, suggesting that the effect is uniquely enabled by the CDW order. Our results reveal a direct interplay between spatial and temporal periodic modulations, establishing a new framework for electronic-phase-selective engineering of the electronic structure through coherent light-matter interactions.

\begin{figure*}[htbp]
	\centering
	\includegraphics[width=16.8cm]{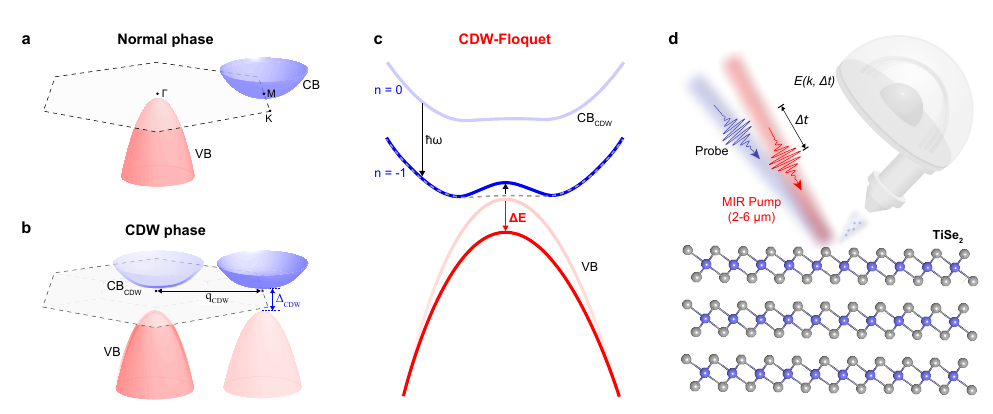}
	\caption*{{\bf Fig.~1 ${\mid}$  Schematics of Floquet engineering in the CDW phase of 1T-TiSe$_2$.}
	 {\bf a,b}, Schematics of electronic structures in the normal phase ($T>T_{\text{CDW}}$, panel a) and in the CDW phase ($T<T_{\text{CDW}}$, panel b), respectively. Red and blue curves represent the VB at the $\Gamma$ point and CB at the M point, which are folded by the CDW wavevector \textbf{q}{$_\text{CDW}$} in the CDW phase. {\bf c}, Schematic illustration of the CB (light blue curve) and VB (light red curve) in the CDW phase, light-induced CB sideband with n = -1 (dashed gray curve), and the renormalized CB sideband (thick blue curve) and VB (thick red curve) upon resonance pumping.
	 {\bf d}, Schematic illustration of TrARPES setup with MIR pumping.}
\label{Fig1}
\end{figure*}


\section*{Floquet band renormalization in the CDW phase upon near-resonance pumping}

1T-TiSe$_2$ is chosen as a model CDW material for studying phase-dependent Floquet engineering due to its layered structure with relatively simple bands. At room temperature, 1T-TiSe$_2$ is characterized by a conduction band minimum (CBM) at the M point and a VBM at the $\Gamma$ point (Fig.~1a). Upon cooling below $T_\text{CDW}$ $\approx$ 200 - 235 K\cite{Waszczak1976PRB,Robert2006NP,jaouen2019phase,chen2016hidden}, the sample undergoes a transition into the commensurate CDW phase with (2$\times$2$\times$2) spatial modulation\cite{rohwer2011nature,li2016nature}. As a result, the conduction band (CB) is folded into the $\Gamma$ point by the CDW vector \textbf{q}{$_\text{CDW}$}, turning its electronic structure into a direct gap semiconductor as schematically illustrated in Fig.~1b, with a reported CDW gap value $\Delta_\text{CDW}$ $\approx$ 220 meV\cite{kidd2002PRL,jaouen2019phase,jaouen2023carrier}. In addition, TiSe$_2$ has also attracted extensive research interests as a candidate material for excitonic insulator\cite{cercellier2007PRL,monney2011exciton,porer2014NM}, chiral CDW\cite{kim2024origin,ishioka2010chiral}. While light-induced phase transition has been revealed upon 800 nm pumping\cite{zhang2021nature,xu2020nature,lanzara2024SA} through photoexcitation and energy transfer, MIR pumping can access a fundamentally distinct regime, where stronger light-fields could favor Floquet engineering, as demonstrated in topological insulators\cite{Gedik2013,Mahmood2025floquet,Bi2Te3PRB2024,Bi2Te3PRL2025} and black phosphorus\cite{zhou2023pseudospin,ZhouBPPRL2023}.

The distinct electronic structure and CDW transition accompanied by transition to a direct gap semiconductor provides an ideal platform for realizing phase-dependent Floquet engineering by coupling the valence band (VB) and CB via light-field dressing (Fig.~1c). For example, when  a resonant pump is applied with photon energy tuned close to the CDW gap, the light-induced n = -1 CB sideband (gray dashed curve) could hybridize with the VB at the $\Gamma$ point (light red curve), leading to a dynamical band renormalization at the touching band edges (thick blue and red curves in Fig.~1c). Crucially, such effect is expected to disappear above $T_{\text{CDW}}$ due to the separation of CB and VB in the momentum space.

\begin{figure*}[htbp]
	\centering
	\includegraphics[width=16.8cm]{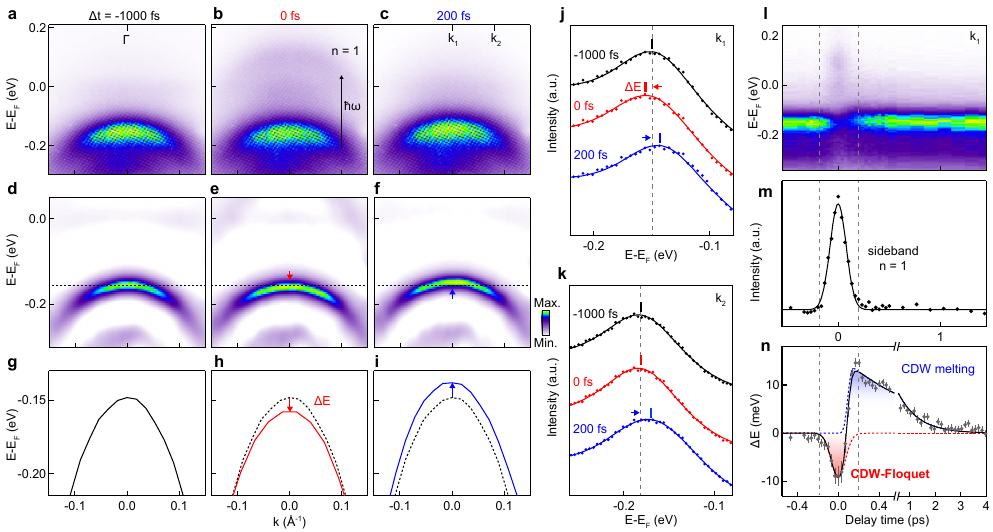}
	\caption*{{\bf Fig.~2 ${\mid}$  Observation of Floquet-induced band renormalization in CDW phase of 1T-TiSe$_2$.}
	 {\bf a-f}, TrARPES dispersion images measured at different delay times along the M-$\Gamma$-M direction by using \textit{p-pol.}~pump ({\bf a-c}), and corresponding second derivative images for direct visualization of the bands ({\bf d-f}). The pump photon energy is 248 meV and the fluence is 2.0 mJ/cm$^{2}$. 
	 {\bf g-i}, Extracted dispersions of VB from ({\bf a-c}) by fitting EDCs. The dotted black curves are extracted dispersion at -1000 fs.
	 {\bf j-k}, EDCs extracted at momenta of $k_1$ and $k_2$ (marked in {\bf c}) at different delay times. Black, red and blue tick marks are the corresponding peak positions.
	 {\bf l}, Intensity map obtained by plotting EDC at momenta $k_1$ as a function of delay time. 
     {\bf m}, TrARPES intensity of n = 1 VB sideband as a function of delay time.
     {\bf n}, Extracted band shift at momenta $k_1$ as a function of delay time. Solid curves are fitting results, while red and blue shaded regions represent the downshift induced by Floquet engineering and upward shift indicating CDW melting.
}\label{Fig2}
\end{figure*}

We start by pumping the sample at wavelength $\lambda$ = 5 $\mu$m with corresponding photon energy of $\hbar\omega$ = 248 meV based on the following considerations. First of all, the photon energy is near resonance with  the CDW gap, which could help to enhance the band hybridization. Secondly, a high pump fluence up to 2.0 mJ/cm$^2$ can be applied on the sample, which corresponds to a peak electric field strength of $E$ = $3.1 \times 10^8$ V/m. Such high peak field and low photon energy is critical for achieving a strong Floquet interaction\cite{Sentef2021}. Figure~2a-c shows a comparison of TrARPES snapshots measured at sample temperature of 80 K for delay time $\Delta t$ = -1000 fs, 0 fs, and 200 fs respectively. At $\Delta t$ = 0, a VB replica displaced from the VB by the pump photon energy  is clearly observed, suggesting light-field dressing of the electronic states. Interestingly, a striking renormalization of the transient electronic structure is observed when comparing data measured at different delay times, which is more clearly resolved in the second-derivative images (Fig.~2d-f, see Methods for details). In particular, the VBM shows a downward shift in energy at $\Delta t$ = 0 (Fig.~2e) and an upward shift at $\Delta t$ = 200 fs (Fig.~2f), which is further supported by the extracted dispersions in Fig.~2g-i. 

We further zoom in to analyze the downward energy shift at $\Delta t$ = 0, which occurs near the VBM as indicated in Fig.~2h. By comparing energy distribution curves (EDCs) at $\Gamma$ ($k_1$) and away from $\Gamma$ ($k_2$) at different delay times, Fig.~2j,k shows a downward energy shift of 8 $\pm$ 3 meV at the $\Gamma$ point (black and red curve in Fig.~2j), which becomes negligible when moving away from the $\Gamma$ point (black and red curves in Fig.~2k). Such momentum-dependent band renormalization is reminiscent of the Floquet band engineering recently observed in black phosphorus\cite{ZhouBPPRL2023}. To further follow the dynamic evolution, Fig.~2l shows the evolution of EDC at the $\Gamma$ point as a function of pump-probe delay time. Here, the intensity near time zero (between the gray dashed lines) corresponds to the transient occupation of the n = 1 VB sideband, whose timescale corresponds to the pump-probe correlation (Fig.~2m). We can also extract the dynamic evolution of the peak shift at $k_1$, which shows first a downshift followed by an upshift in Fig.~2n. By fitting the data using two dynamic processes, we can decompose it into a fast downshift (indicated by red curve) and a slow upshift (indicated by blue curve). Interestingly, the instantaneous downshift shows similar dynamics to the n = 1 VB sideband (Fig.~2m), suggesting that it is likely induced by the light field coherently. In contrast, the upward energy shift of the VBM (blue dashed curve in Fig.~2n) has a much longer timescale of 1.4 ps, in line with CDW melting upon pumping at 800 mm\cite{rohwer2011nature,mathias2016self,hedayat2019PRR,zhou2025CPL}. Therefore, the overall band renormalization can be described by an instantaneous downshift suggesting Floquet engineering, followed by upward shift on picosecond timescale indicating CDW melting.

\section*{Floquet engineering from near-resonance to off-resonance pumping}

\begin{figure*}[htbp]
	\centering
	\includegraphics[width=16.8cm]{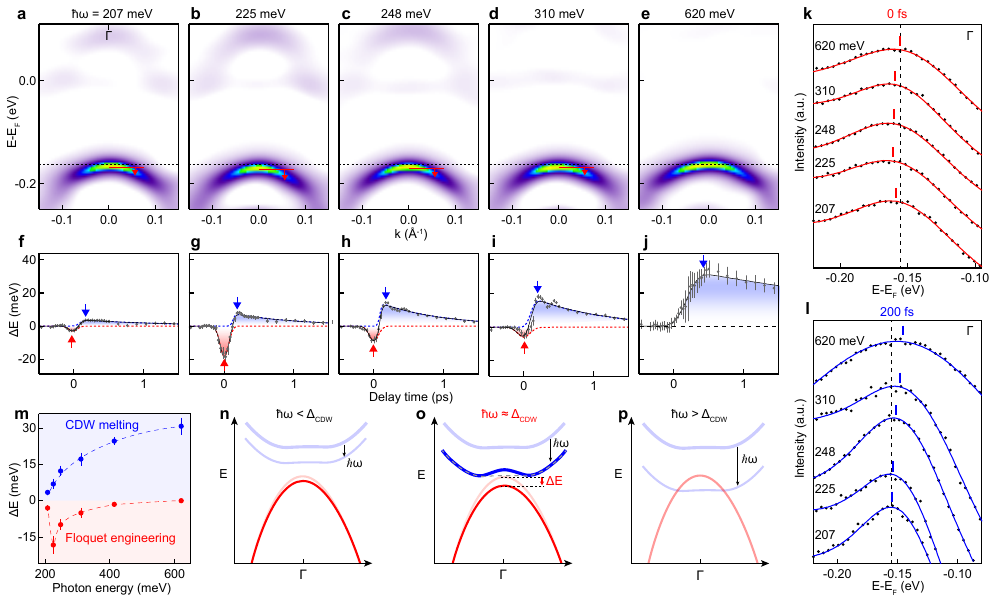}
	\caption*{{\bf Fig.~3 ${\mid}$ Evolution of Floquet-induced band renormalization when tuning the pump photon energy from resonance to off-resonance.}
     {\bf a-e}, Second derivative TrARPES dispersion images measured at $\Delta t$ = 0 when tuning the pump photon energy from 207 to 620 meV. The pump fluence is fixed at 2.0 mJ/cm$^2$.
	 {\bf f-j}, Time-resolved energy shift of n = 0 VB extracted at the $\Gamma$ point at different pump photon energies.
     {\bf k-l}, EDCs at the $\Gamma$ point at different pump photon energies for delay time of 0 {(\bf k)} and 200 fs {(\bf l)}. The red and blue tick marks are the corresponding peak positions.
	 {\bf m}, Extracted energy shift at the $\Gamma$ point for different pump photon energies, where the error bar is determined from the curve fitting results in {\bf f-j}. 
	 {\bf n-p}, Schematic summary of transient electronic structures upon below-gap pumping, resonant pumping and above-gap pumping.
	}\label{Fig3}
\end{figure*}

To investigate whether the light-induced band renormalization is enhanced by resonance pumping with the CDW gap, we systematically tune the pump photon energy from 207 meV to 620 meV. Figure~3a-e shows dispersion images measured at $\Delta t$ = 0 for different pump photon energies. A clear n = 1 VB sideband is observed at lower photon energies (Fig.~3a-c). The corresponding temporal evolution of the energy shift of VBM at the $\Gamma$ point is plotted in Fig.~3f-j, revealing two distinct processes: an instaneous downward energy shift (red shading) attributed to Floquet-induced band renormalization, and a slower upward shift (blue shading) due to CDW gap melting. 
At low pump photon energies (Fig.~3f-i), both effects are present, with the Floquet-induced downward shift (red arrows) maximized upon pumping at 225 meV (supported by EDC analysis shown in Fig.~3k), suggesting an enhanced Floquet engineering when pumping resonantly with the CDW gap. In contrast, the CDW-melting-induced upward shift (blue arrows) grows monotonically with increasing photon energy (see EDC analysis in Fig.~3l). At larger pump photon energy, for example, 620 meV, only the upward shift remains (Fig.~3j), indicating the absence of Floquet renormalization when detuned from resonance pumping.  Figure 3m shows the extracted VBM energy shift obtained by fitting data using two dynamic processes (red and blue curves in Fig.~3f-j, representing energy shift induced by Floquet engineering and CDW melting respectively), where a resonance enhancement is clearly observed for the Floquet-induced energy shift (red curve).

The pump-photon-energy-dependent Floquet-induced renormalizations from below-gap to above-gap pumping are schematically summarized in Fig.~3n-p. For below-gap pumping (Fig.~3n), the n = -1 CB sideband is well separated from the VB, resulting in minimal band renormalization. As the pump photon energy increases and approaches resonance (Fig.~3o), the CB sideband (blue curve) becomes closer to the VBM, leading to pronounced band renormalization which is evidenced by a maximum downward shift of the VBM (indicated by the red arrow). At even higher energy (detuned from resonance pumping), the Floquet-induced renormalization effect becomes negligible (Fig.~3p). These observations point to a resonance-enhanced light-matter interaction, which strongly amplifies the Floquet-induced band renormalization at the VBM.

\section*{Floquet engineering observed only in the CDW phase}

\begin{figure*}[htbp]
	\centering	
	\includegraphics[width=16.8cm]{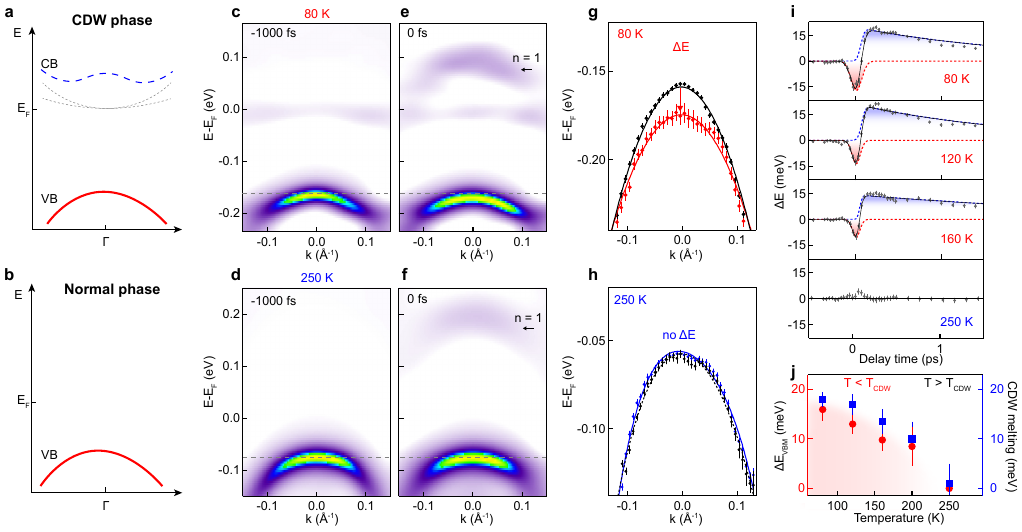}
  \caption*{{\bf Fig.~4 ${\mid}$ Interplay between the CDW order and Floquet band renormalization.}
   {\bf a,b}, Schematic dispersion of TiSe$_{2}$ in the CDW phase ({\bf a}) and the normal phase ({\bf b}). 
   {\bf c,d}, Second derivative dispersion images measured at $\Delta t$ = -1000 fs at 80 K ({\bf c}) and at 250 K ({\bf d}). 
   {\bf e,f}, Second derivative dispersion images measured at $\Delta t$ = 0 fs at 80 K ({\bf e}) and at 250 K ({\bf f}). The pump photon energy is 248 meV with \textit{p-pol.} and a pump fluence of 2 mJ/cm$^{2}$. 
   {\bf g,h},  Extracted dispersions of VB at 80 K ({\bf g}) and 250 K ({\bf h}) by fitting EDCs from {\bf c,e} and {\bf d,f} respectively. 
   {\bf i}, Time-resolved energy shift of VBM at the $\Gamma$ point measured at different sample temperatures. 
   {\bf j}, Extracted Floquet-induced energy shifts (red circles) and CDW-melting-induced energy shifts (blue squares) at the $\Gamma$ points as a function of temperature. 
    }\label{Fig4}
\end{figure*}

To further reveal the possible interplay between Floquet engineering and CDW phase transition, we show in Fig.~4 direct comparison of data measured at 80 K (in the CDW phase) and 250 K (above $T_{\text{CDW}}$) with pump photon energy $\hbar \omega $ = 248 meV. In the CDW phase ($T < T_{\text{CDW}}$), the CB at the M point is folded to the $\Gamma$ point, forming a CDW-folded CB (Fig.~4a), while such folded CB is absent at the $\Gamma$ point in normal phase (Fig.~4b). Such a difference is clearly visualized in Fig.~4c,d, where the CDW-folded CB is only observed in the CDW phase (Fig.~4c) as compared with normal phase (Fig.~4d). When applying near-resonance pumping, a light-induced n = 1 VB sideband is observed at $\Delta t$ = 0 fs for both CDW and normal phases as shown in Fig.~4e,f. However, the Floquet-induced downward energy shift is observed only in the CDW phase (Fig.~4e), which is clearly evidenced by comparing the extracted dispersions in Fig.~4g,h, suggesting that the Floquet induced energy shift occurs only in the CDW phase.

The phase-dependent Floquet engineering is further supported by more detailed temperature-dependent measurements. By varying the sample temperature from 80 K to 250 K, a clear suppression of the light-induced band renormalization is observed upon increasing temperature, with the effect vanishing completely in the normal phase (Fig.~4i). 
The Floquet-induced energy shift (red circles in Fig.~4j) shows a similar temperature evolution with the CDW-melting-induced energy shift (blue squares in Fig.~4j), and they are both observed only below the CDW transition temperature. Such direct correspondence provides compelling evidence for the essential role of CDW order in enabling Floquet engineering, highlighting the strong interplay between the spontaneous symmetry breaking and light-field driven band engineering.

\section*{Theoretical modeling}

\begin{figure*}[htbp]
	\centering	
	\includegraphics[width=16.8cm]{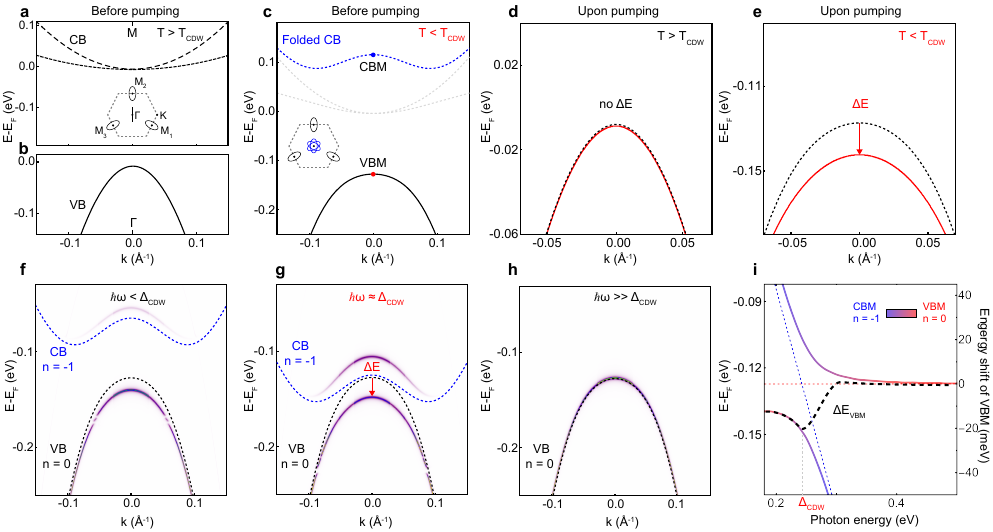}
    \caption*{{\bf Fig.~5 ${\mid}$ Numerical results from the  four-band model for 1T-TiSe$_2$.} 
	{\bf a,b}, The dispersion of the equilibrium electronic structure in the normal phase ($T>T_{\text{CDW}}$) along the $\Gamma$-M$_2$ direction. The CBs ({\bf a}) are from three M points and the VB ({\bf b}) is at $\Gamma$ point. The dashed line in ({\bf a}) represents a doubly degenerate CB. The inset is the Brillouin zone with three CB pockets (black ellipses) at three M points.
	{\bf c}, Dispersion of the equilibrium VB (black curve) and folded CBs (dashed curves) in the CDW phase ($T<T_{\text{CDW}}$). The inset is the Brillouin zone with three CB pockets folded to the $\Gamma$ point.
	{\bf d,e}, Dispersion of the VB upon pumping in the normal phase ({\bf d}) and CDW phase ({\bf e}) compared to the equilibrium VB (black dashed curve). The pump photon energy is 220 meV.
	{\bf f-h}, TrARPES intensity with the photon energy smaller than ({\bf f}, 180 meV), almost equal to ({\bf g}, 240 meV), and much larger than ({\bf g}, 400 meV) $\Delta_{\text{CDW}}$. The red arrow in ({\bf g}) indicates the energy shift of the VBM. The black (blue) dashed curve is the equilibrium VB (n = -1 replica of the CB) in the CDW phase. 
	{\bf i}, Energies of the Floquet electronic states (solid curves) as a function of the pump photon energy after hybridizing the VB and the n = -1 replica of the CB (thin dashed lines). The black dashed curve is a schematic interpolation between them.
    }\label{Fig5}
\end{figure*}

To understand the key role played by CDW order in the Floquet engineering, we employ a four-band  model for the low energy electrons coupled with CDW in 1T-TiSe$_2$ (see Methods for more details). 
It includes one highest VB at the $\Gamma$ point and the three lowest CBs, one from each M point, as shown in Fig.~5a,b.
At temperatures below $T_{\text{CDW}}$, 
the three-component CDW order parameter $\Delta=(\Delta_1,\Delta_2,\Delta_3)$ is nonzero, 
which may arise from the lattice displacements $X(\vec{q}_i)$ at the corresponding wave vectors $\vec{q}_{1}$, $\vec{q}_{2}$, $\vec{q}_{3}$ connecting the $\Gamma$ point and the three M points, or from excitonic order\cite{kidd2002PRL,cercellier2007PRL,monney2011exciton,porer2014NM}.
The order parameter folds the three M points to the $\Gamma$ point and hybridizes the VB and the three CBs, so that the system may be viewed as a direct gap semiconductor at the $\Gamma$ point shown in Fig.~5c. 
Among the three hybridized CBs in the CDW phase, only the highest one shown as blue dashed line  in Fig.~5c has a substantial hybridization with the VB, therefore the energy gap between these two bands is defined as the CDW gap, as in the literature\cite{watson2019orbital,jaouen2019phase,watson2020strong,jaouen2023carrier}.

Given that the $C_3$ rotational symmetry is preserved in the CDW phase\cite{Huang_2021,Zhang2022prb}, 
the order parameter could be expanded as $\Delta=\Delta_0+\Delta_1$, where the $\Delta_0$ term is independent on the momentum $k$ and respects the inversion symmetry, while the term $\Delta_1 \propto k$ breaks the inversion symmetry\cite{Zhang2022prb} (see Methods for more details). 
In the normal phase, one has $\Delta=0$ so that the pump light has zero matrix elements between the VB and CB, which are located at different momentum positions.
Therefore, only intraband  Floquet effects such as the replicas and ponderomotive shifts appear upon pumping, while no interband Floquet effects occur (see Fig.~5d). In the CDW phase, the situation is completely different  because the  CDW order parameter leads to nonzero optical matrix elements between the VB and CB.
As shown in Fig.~5e, interband Floquet  effect renormalizes the original VB (black dashed line)  to the red curve upon below-gap pumping.
Note that because of the high in-plane dielectric screening (see Methods for details), the in-plane electric field of the \textit{p-pol.}~pump is greatly suppressed in the sample, and the dominant interband Floquet effect comes from its out-of-plane electric field via the
$\Delta_1 \propto k_z$ component of the order parameter, consistent with the polarization dependent TrARPES experimental results (see Extended Data Fig.~1 and Fig.~2).

Figures 5f-h show the numerical results of the TrARPES intensity upon below-gap ($\hbar\omega<\Delta_{\text{CDW}}$, see panel f), near-resonance  ($\hbar\omega\approx\Delta_{\text{CDW}}$, panel g), and far-above-gap pumping ($\hbar\omega\gg\Delta_{\text{CDW}}$, panel h), respectively. 
Consistent with the experimental observation, the VBM downward shift is small for below-gap pumping (Fig.~5f), increases and reaches maximum as the pump photon energy approaches interband resonance (Fig.~5g), and becomes negligible for far above-gap pumping (Fig.~5h). In Fig.~5i, we plot the energies of the VBM (red dotted line), the CBM for n = -1 sideband (blue dotted line), and the energies of the Floquet states after hybridization (solid blue-to-red curves) as a function of the pump photon energy. Because of the strong hybridization between the CBM (n = -1) and VBM (n = 0) upon near-resonance pumping, the experimentally observed peak position in Fig.~3m is likely contributed by both bands, and could be understood as the schematic interpolation curve in Fig.~5i (black dashed curve, indicated by $\Delta {\text E_{VBM}}$). Note that because the dominant interband optical matrix element is between the VBM and the CBM, the Floquet engineering effect is maximized when the pump photon approaches this CDW gap, consistent with the experiment observation (see red curve in Fig.~3m).

\section*{Conclusion and outlook}

Our results unambiguously demonstrate Floquet band renormalization in the CDW material 1T-TiSe$_2$. This effect arises from the dynamical coupling between Floquet-modulated, CDW-folded CB and the VB. The observation of resonant Floquet effects only within the CDW phase, and their dependence on pump photon energy, highlights the crucial role of the CDW order in enabling the Floquet engineering. These findings uncover a previously unexplored interplay between spatial symmetry breaking and time-periodic driving, establishing a new regime for electronic-phase-dependent Floquet engineering. Our work opens a pathway toward realizing Floquet engineering in a broader class of quantum materials with electronic phase transitions including charge/spin density waves, excitonic insulators and superconductors.

\begin{methods}

\subsection{Single crystal growth.}
 1T-TiSe$_2$ single crystals were grown by the chemical vapour transport method. Polycrystalline 1T-TiSe$_2$ were synthesized
 by heating the high-purity stoichiometric mixture of Ti powder (99.9$\%$, Alfa Aesar) and Se ingot (99.999$\%$, Macklin) at 900$^\circ$C in a vacuum-sealed silica ampoule for three days. Using I$_2$ (5 mg$\cdot$ml$^{-1}$) as transfer agent, the polycrystalline 1T-TiSe$_2$ were then transferred from high temperature (670$^\circ$C) area to low temperature (600$^\circ$C) area to achieve recrystallization. High-quality 1T-TiSe$_2$ single crystals were obtained over a period of 30 days.
 
\subsection{TrARPES measurements.}
 TrARPES measurements were performed in our home laboratory based on a Ti:sapphire laser amplifier with center wavelength of 800 nm and repetition rate of 10 kHz. The laser is split into two beams, a major part (90\%) of the laser beam is used to drive an optical parametric amplifier (OPA) and non-collinear differential frequency
 generation (NDFG) to generate MIR pump. The probe beam with photon energy of 6.2 eV is generated by a three-step  fourth harmonics generation process using BBO crystals. The samples were cleaved in ultrahigh vacuum chamber with a base pressure better than 1 $\times$ 10$^{-10}$ Torr and measured at a temperature of 80 K unless otherwise specified.

 A conventional approach to directly visualize the dispersion is to take the second derivative of the dispersion images, for example, Fig.~2d-f. The second derivative (or 2D curvature) of the ARPES intensity $I_{\text{ARPES}}$ is
 \begin{equation}
 \nabla^{2}I_{\text{ARPES}}=\frac{\partial^{2}I_{\text{ARPES}}}{\partial x^{2}}+\frac{\partial^{2}I_{\text{ARPES}}}{\partial y^{2}}
 \end{equation}
 
By applying a coordinate transformation to $x$ and $y$ to account for the unit difference of $k$ and $E$, the equation takes the form

\begin{equation}
C(x,y)\approx\frac{[1+C_{x}(\frac{\partial f}{\partial x})^{2}]C_{y}\frac{\partial^{2}f}{\partial y^{2}}-2C_{x}C_{y}\frac{\partial f}{\partial x}\frac{\partial f}{\partial y}\frac{\partial^{2} f}{\partial x\partial y}+[1+C_{y}(\frac{\partial f}{\partial y})^{2}]C_{x}\frac{\partial^{2}f}{\partial x^{2}}}{[1+C_{x}(\frac{\partial f}{\partial x})^{2}+C_{y}(\frac{\partial f}{\partial y})^{2}]^{3/2}}
\end{equation}

Such approach has been widely utilized in the ARPES data analysis for a direct visualization of electronic dispersions.

\subsection{Effective model of 1T-TiSe$_2$.}
To understand the experiment, we employ a four-band effective model for 1T-TiSe$_2$. 
The mean field Hamiltonian  is
\begin{equation}\label{H0}
\begin{aligned}
&{H}=
\begin{pmatrix}
	{v}_{k}^\dagger
	&
	{c}_{1,k}^\dagger
	&
	{c}_{2, k}^\dagger
	&
	{c}_{3, k}^\dagger
\end{pmatrix}
		\begin{pmatrix}
			\varepsilon_{k}^v
			& \Delta_1 & \Delta_2 & \Delta_3
			\\
			\Delta_1^{\ast} & \varepsilon_{1,k}^c & 0 & 0
			\\
            \Delta_2^{\ast} & 0 & \varepsilon_{2,k}^c & 0 
   			\\
           \Delta_3^{\ast} & 0 & 0 & \varepsilon_{3,k}^c
		\end{pmatrix}
		\begin{pmatrix}
		{v}_{k}
			\\
		{c}_{1,k}
       \\
       	{c}_{2, k}
       \\		
       {c}_{3, k}
		\end{pmatrix}
\,.
\end{aligned}
\end{equation}
It consists of one VB at the $\Gamma$ point and three CBs, one from each M point, as shown in the inset of Fig.~5a. 
Here $\varepsilon_{k}^v = a(k_x^2 + k_y^2) + c k_z^2 + \epsilon_1$ is the kinetic energy of the VB with corresponding annihilation (creation) operators $v_k$ ($v^\dagger_k$) and $k_x$ ($k_y$) is along the $\Gamma \mbox{-} K$ ($\Gamma \mbox{-} M_2$) direction, as shown in the inset of Fig.~5a. 
The kinetic energy of the CB around the $M_2$ point is
$\varepsilon_{2,k}^c = A k_x^2 + B k_y^2 + C k_z^2 + \epsilon_2$ with the corresponding annihilation (creation) operators $c_{2,k}$ ($c^\dagger_{2,k}$), 
where $k$ is the electronic momentum defined relative to the $M_2$  point in the Brillouin  zone.
Because of the three-fold rotational symmetry ($\mathbf{C}_3$), the kinetic energy of the CBs at the $M_1$  and $M_3$  points are given by: $\varepsilon_{1,k}^c=\varepsilon_{2,\hat{C}k}^c,
\varepsilon_{3,k}^c=\varepsilon_{2,\hat{C}^{-1}k}^c$\cite{Huang_2021}, 
where $\hat{C}$ represents the 120$^\circ$-rotational matrix  for vectors. 
With this definition, the three conduction bands are already shifted to the $\Gamma$ point in \equa{H0}.
Following Ref.\cite{kim2024origin}, the parameters employed are: 
$a = -20.0$ eV$\cdot$\AA$^2$, $c= 0.04$ eV$\cdot$\AA$^2$, $A = 6.3 $ eV$\cdot$\AA$^2$, $B = 1.5$ eV$\cdot$\AA$^2$, $C = 0.24$ eV$\cdot$\AA$^2$, $\epsilon_1 = -0.008$ eV, $\epsilon_2 = -0.004$ eV, corresponding to a semiconductor in the normal phase.

The complex CDW order parameter $\Delta=(\Delta_1,\Delta_2,\Delta_3)$ has three components, corresponding to the mean field hybridization  matrix elements between the valence band and the three conduction bands, respectively.
In the case of CDW induced by pure lattice displacement, the three order parameter components arise from the lattice displacements $X(q_i)$ at the corresponding wave vectors $\vec{q}_{1}, \vec{q}_{2}, \vec{q}_{3}$ connecting the $\Gamma$ point and the three M points. 
The order parameter may also have contributions from the excitonic order\cite{cercellier2007PRL,monney2011exciton,porer2014NM}. In either case, the mean field Hamiltonian could all be described by the same form in \equa{H0}. 
In principle, $\Delta$ is a function of $k$. If expanded to linear order in $k$, the form of $\Delta_k$ that preserves the $C_3$ rotation symmetry of the system reads
\begin{align}\label{eqn:delta}
\Delta=(\Delta_0,\, \Delta_0,\, \Delta_0)
+ \lambda   (\hat{C} \vec{q},\,  \vec{q},\, \hat{C}^{-1} \vec{q})  \cdot  \vec{k}
\end{align}
where $\vec{q}$ is an arbitrary vector.
The constant terms $(\Delta_0,\, \Delta_0,\, \Delta_0)$
also preserve the spatial-inversion symmetry since all the four bands have even parity under inversion\cite{Zhang2022prb}. 
Although the linear-in-$k$ term breaks the inversion symmetry\cite{Zhang2022prb}, we still keep this term, because it is not clear from previous literature whether the CDW spontaneously breaks the inversion symmetry or not. 
In the CDW phase, the order parameter hybridizes the valence band and conduction bands and pushes the valence band downward. At the $\Gamma$ point, only one of the linearly combined conduction band $c_0=\frac{1}{\sqrt{3}} (c_{1,0}+c_{2,0}+c_{3,0})$ hybridizes with the valence band and is pushed up to the highest one, while the other two are decoupled from the valence band, see Fig.~5c. Slightly away from the $\Gamma$ point, the picture remains roughly the same.

The pump light enters the Hamiltonian by the standard Peierls substitution: $k \rightarrow k + A(t)$, where $A(t) = A_0 \mathrm{e}^{-\mathrm{i}\omega t} + \text{c.c.}$ is the uniform vector potential oscillating at  frequency $\omega$. 
The electric field is $E(t) = -\partial_t A(t)$ and $E_0 = \mathrm{i} A_0 \omega$. 
The single-electron properties can be calculated from the non-equilibrium Green function
\begin{equation}\label{eqn:dyson}
	\begin{aligned}
		{G} & =\left(\begin{array}{cc}
			G^R & G^K \\
			0 & G^A
		\end{array}\right)
		=\left({G}_0^{-1}-{\Sigma}\right)^{-1}
	\end{aligned}
\end{equation}
perturbatively in both the pump field and any interactions\cite{xiao2025arXiv}.
Here, $G_0$ is the equilibrium bare Green function with the correction from the mean field order parameter, and
the self-energy $\Sigma$ reads
\begin{equation}\label{eqn:self_energy}
	\begin{aligned}
		\Sigma  ={\Gamma}(-\omega){G}_{0}(\boldsymbol{\omega}+\omega){\Gamma}(\omega)+{\Gamma}(\omega){G}_{0}(\boldsymbol{\omega}-\omega){\Gamma}(-\omega)
	\end{aligned}
\end{equation}
to the lowest-order in the pump field, where $\Gamma(-\omega) = \Gamma(\omega) = A_0 \cdot \partial_k H(k)$ is the bare light-matter vertex.
In principle, there could be fluctuation corrections to the self-energy from the collective modes such as the CDW phase and amplitude modes, which we leave for future theoretical study.
Plugging \equa{eqn:self_energy} into \equa{eqn:dyson}, one obtains the lesser Green function  $G^<=(G^K+G^A-G^R)/2$ which renders the Floquet-Bloch bands  and the TrARPES intensity in Fig.~5.  

For the numerical plots in Fig.~5 along the $\Gamma\mbox{-}M_2$ direction ($k= k_y$  and $k_z = 0$), 
we set  $E_0 = 7 \times 10^5 \unit{V/cm}$ and $\Delta_0 = 0.07 \unit{eV}$. 
We have also set $\lambda = i$  due to time-reversal symmetry and 
$\vec{q}=(0,\, 0,\,  0.3  \unit{eV \cdot \AA})$. Since our investigation is restricted to the $\Gamma$ point and its immediate vicinity, in the normal phase, we set the ARPES probe matrix elements of conduction bands to zero, and that of the valence band as a momentum-independent constant.

\subsection{Transmission of the pump into the sample.}

Note that the interband hybridization is experimentally observed for  \textit{p-pol.} pump light only (see Extended Data Fig.~1) but not for \textit{s-pol.} pump, and is also independent of measurement direction of the crystal orientation (see Extended Data Fig.~2).
Part of the reason is that because of the high dielectric along the in-plane direction (along the surface of the sample),  the in-plane electric field of the pump light could hardly transmit to inside the sample. 
Therefore, it is important to perform an electromagnetic (EM) calculation of the
the strength of the pump field inside the sample, which could be obtained from its dielectric tensor $(\epsilon_x, \epsilon_z)$ where $\epsilon_x$ ($\epsilon_z$) is the in-plane (out-of-plane) dielectric function.

For \textit{p-pol.} pump with the wave vector $\vec{k}_0=(k_{0x}, k_{0z})=k_0 (\sin \theta,\, \cos\theta)$, the transmitted light inside the sample has the wave vector
$(k_{x}, k_{z})=
\left( k_{0x},\, \sqrt{\epsilon_x (\omega^2/c^2-k_{0x}^2/\epsilon_z)} 
\right)$. 
From the relation between the  $\vec{D}=- \frac{c}{\omega} \vec{k} \times \vec{B}=\hat{\epsilon} \vec{E}$, one obtains the pump electric field inside the sample:
\begin{align}
(E_x, E_z)=B_0 T_p \frac{1}{k_0} \left( \frac{k_z}{\epsilon_x},\,  -\frac{k_x}{\epsilon_z} \right)
=B_0  T_p \frac{1}{k_0} \left( \frac{\sqrt{\epsilon_x (\omega^2/c^2-k_{0x}^2/\epsilon_z)}}{\epsilon_x},\,  -\frac{k_{0x}}{\epsilon_z} \right)
\,.
\end{align}
where $T_p=2k_{0z}/(k_{0z}+\frac{k_z}{\epsilon_x})$ is the Fresnel  transmission coefficient and  $B_0 =E_0= 7 \times 10^5 \unit{V/cm}$ is the amplitude of the magnetic (electric) field in Gaussian unit of the pump pulse in vacuum.
Because of the layered structure of 1T-TiSe$_2$, its dielectric is highly anisotropic.
From optical measurements\cite{Wilson.1978}, the normal incidence reflectivity is about $|R|^2=0.55$ in the frequency range of our pump ($2000$ - $6000 \unit{cm^{-1}}$), which corresponds to an in-plane dielectric $\epsilon_x \approx 46$. 
The out-of-plane dielectric is unknown which we set as $\epsilon_z=1$
 considering its layered structure. With these numbers and the incident angle $\theta=53^\circ$, we obtain $(E_x,\, E_z)=(0.15,\, -1.4)E_0$ inside the sample, meaning that the in plane electric field is much weaker than the out-of-plane one.
 In Fig.~5, the Floquet engineering effects come mainly from $E_z$ in the sample.

For \textit{s-pol.} pump, similar EM analysis gives the in-plane electric field:
\begin{align}
E_x=E_0 T_s = E_0 \frac{2k_{0z}}{k_{0z}+k_z} \,.
\end{align}
With $\epsilon_x \approx 46$, one obtains $E_x \approx 0.26  E_0$. This explains why \textit{s-pol.} does not result in an obvious Floquet engineering effect in the experiment. Another reason is that, the CDW hybridized  $c_0=\frac{1}{\sqrt{3}} (c_{1,0}+c_{2,0}+c_{3,0})$ has zero  in-plane optical matrix element with the valence band: the three Peierls terms $\lambda   (\hat{C} \vec{q},\,  \vec{q},\, \hat{C}^{-1} \vec{q})_x  \cdot  A_x$ from \equa{eqn:delta} add up to zero for this band.

\end{methods}

\section*{References}
\bibliography{reference}

\begin{thebibliography}{10}
\expandafter\ifx\csname url\endcsname\relax
  \def\url#1{\texttt{#1}}\fi
\expandafter\ifx\csname urlprefix\endcsname\relax\def\urlprefix{URL }\fi
\providecommand{\bibinfo}[2]{#2}
\providecommand{\eprint}[2][]{\url{#2}}

\bibitem{oka2019floquet}
\bibinfo{author}{Oka, T.} \& \bibinfo{author}{Kitamura, S.}
\newblock \bibinfo{title}{{Floquet engineering of quantum materials}}.
\newblock \emph{\bibinfo{journal}{Annu. Rev. Condens. Matter Phys.}}
  \textbf{\bibinfo{volume}{10}}, \bibinfo{pages}{387} (\bibinfo{year}{2019}).

\bibitem{rudner2020NRP}
\bibinfo{author}{Rudner, M.~S.} \& \bibinfo{author}{Lindner, N.~H.}
\newblock \bibinfo{title}{{Band structure engineering and non-equilibrium
  dynamics in Floquet topological insulators}}.
\newblock \emph{\bibinfo{journal}{Nat. Rev. Phys.}}
  \textbf{\bibinfo{volume}{2}}, \bibinfo{pages}{229} (\bibinfo{year}{2020}).

\bibitem{ZhouNRP2021}
\bibinfo{author}{Bao, C.}, \bibinfo{author}{Tang, P.}, \bibinfo{author}{Sun,
  D.} \& \bibinfo{author}{Zhou, S.}
\newblock \bibinfo{title}{{Light-induced emergent phenomena in 2D materials and
  topological materials}}.
\newblock \emph{\bibinfo{journal}{Nat. Rev. Phys.}}
  \textbf{\bibinfo{volume}{4}}, \bibinfo{pages}{33} (\bibinfo{year}{2021}).

\bibitem{Sentef2021}
\bibinfo{author}{de~la Torre, A.} \emph{et~al.}
\newblock \bibinfo{title}{{Colloquium: Nonthermal pathways to ultrafast control
  in quantum materials}}.
\newblock \emph{\bibinfo{journal}{Rev. Mod. Phys.}}
  \textbf{\bibinfo{volume}{93}}, \bibinfo{pages}{041002}
  (\bibinfo{year}{2021}).

\bibitem{oka2009PRB}
\bibinfo{author}{Oka, T.} \& \bibinfo{author}{Aoki, H.}
\newblock \bibinfo{title}{{Photovoltaic Hall effect in graphene}}.
\newblock \emph{\bibinfo{journal}{Phys. Rev. B}} \textbf{\bibinfo{volume}{79}},
  \bibinfo{pages}{081406} (\bibinfo{year}{2009}).

\bibitem{demler2011PRB}
\bibinfo{author}{Kitagawa, T.}, \bibinfo{author}{Oka, T.},
  \bibinfo{author}{Brataas, A.}, \bibinfo{author}{Fu, L.} \&
  \bibinfo{author}{Demler, E.}
\newblock \bibinfo{title}{{Transport properties of nonequilibrium systems under
  the application of light: Photoinduced quantum Hall insulators without Landau
  levels}}.
\newblock \emph{\bibinfo{journal}{Phys. Rev. B}} \textbf{\bibinfo{volume}{84}},
  \bibinfo{pages}{235108} (\bibinfo{year}{2011}).

\bibitem{lindner2011NP}
\bibinfo{author}{Lindner, N.~H.}, \bibinfo{author}{Refael, G.} \&
  \bibinfo{author}{Galitski, V.}
\newblock \bibinfo{title}{{Floquet topological insulator in semiconductor
  quantum wells}}.
\newblock \emph{\bibinfo{journal}{Nat. Phys.}} \textbf{\bibinfo{volume}{7}},
  \bibinfo{pages}{490} (\bibinfo{year}{2011}).

\bibitem{Podolsky2013PRL}
\bibinfo{author}{Katan, Y.~T.} \& \bibinfo{author}{Podolsky, D.}
\newblock \bibinfo{title}{{Modulated Floquet topological insulators}}.
\newblock \emph{\bibinfo{journal}{Phys. Rev. Lett.}}
  \textbf{\bibinfo{volume}{110}}, \bibinfo{pages}{016802}
  (\bibinfo{year}{2013}).

\bibitem{claassen2016NC}
\bibinfo{author}{Claassen, M.}, \bibinfo{author}{Jia, C.},
  \bibinfo{author}{Moritz, B.} \& \bibinfo{author}{Devereaux, T.~P.}
\newblock \bibinfo{title}{{All-optical materials design of chiral edge modes in
  transition-metal dichalcogenides}}.
\newblock \emph{\bibinfo{journal}{Nat. Commun.}} \textbf{\bibinfo{volume}{7}},
  \bibinfo{pages}{13074} (\bibinfo{year}{2016}).

\bibitem{yan2016PRL}
\bibinfo{author}{Yan, Z.} \& \bibinfo{author}{Wang, Z.}
\newblock \bibinfo{title}{{Tunable Weyl points in periodically driven nodal
  line semimetals}}.
\newblock \emph{\bibinfo{journal}{Phys. Rev. Lett.}}
  \textbf{\bibinfo{volume}{117}}, \bibinfo{pages}{087402}
  (\bibinfo{year}{2016}).

\bibitem{zhang2016PRB}
\bibinfo{author}{Zhang, X.-X.}, \bibinfo{author}{Ong, T.~T.} \&
  \bibinfo{author}{Nagaosa, N.}
\newblock \bibinfo{title}{{Theory of photoinduced Floquet Weyl semimetal
  phases}}.
\newblock \emph{\bibinfo{journal}{Phys. Rev. B}} \textbf{\bibinfo{volume}{94}},
  \bibinfo{pages}{235137} (\bibinfo{year}{2016}).

\bibitem{Rubio2017NC}
\bibinfo{author}{H{\"u}bener, H.}, \bibinfo{author}{Sentef, M.~A.},
  \bibinfo{author}{De~Giovannini, U.}, \bibinfo{author}{Kemper, A.~F.} \&
  \bibinfo{author}{Rubio, A.}
\newblock \bibinfo{title}{{Creating stable Floquet--Weyl semimetals by
  laser-driving of 3D Dirac materials}}.
\newblock \emph{\bibinfo{journal}{Nat. Commun.}} \textbf{\bibinfo{volume}{8}},
  \bibinfo{pages}{13940} (\bibinfo{year}{2017}).

\bibitem{Gedik2013}
\bibinfo{author}{Wang, Y.}, \bibinfo{author}{Steinberg, H.},
  \bibinfo{author}{Jarillo-Herrero, P.} \& \bibinfo{author}{Gedik, N.}
\newblock \bibinfo{title}{{Observation of Floquet-Bloch states on the surface
  of a topological insulator}}.
\newblock \emph{\bibinfo{journal}{Science}} \textbf{\bibinfo{volume}{342}},
  \bibinfo{pages}{453} (\bibinfo{year}{2013}).

\bibitem{Gedik2016}
\bibinfo{author}{Mahmood, F.} \emph{et~al.}
\newblock \bibinfo{title}{{Selective scattering between Floquet-Bloch and
  Volkov states in a topological insulator}}.
\newblock \emph{\bibinfo{journal}{Nat. Phys.}} \textbf{\bibinfo{volume}{12}},
  \bibinfo{pages}{306} (\bibinfo{year}{2016}).

\bibitem{zhou2023pseudospin}
\bibinfo{author}{Zhou, S.} \emph{et~al.}
\newblock \bibinfo{title}{{Pseudospin-selective Floquet band engineering in
  black phosphorus}}.
\newblock \emph{\bibinfo{journal}{Nature}} \textbf{\bibinfo{volume}{614}},
  \bibinfo{pages}{75} (\bibinfo{year}{2023}).

\bibitem{HuberNature2023}
\bibinfo{author}{Ito, S.} \emph{et~al.}
\newblock \bibinfo{title}{{Build-up and dephasing of Floquet--Bloch bands on
  subcycle timescales}}.
\newblock \emph{\bibinfo{journal}{Nature}} \textbf{\bibinfo{volume}{616}},
  \bibinfo{pages}{696} (\bibinfo{year}{2023}).

\bibitem{Mahmood2025floquet}
\bibinfo{author}{Bielinski, N.} \emph{et~al.}
\newblock \bibinfo{title}{{Floquet--Bloch manipulation of the Dirac gap in a
  topological antiferromagnet}}.
\newblock \emph{\bibinfo{journal}{Nat. Phys.}} \textbf{\bibinfo{volume}{21}},
  \bibinfo{pages}{458} (\bibinfo{year}{2025}).

\bibitem{stfan2025graphene}
\bibinfo{author}{Merboldt, M.} \emph{et~al.}
\newblock \bibinfo{title}{{Observation of Floquet states in graphene}}.
\newblock \emph{\bibinfo{journal}{Nat. Phys.}} \textbf{\bibinfo{volume}{21}},
  \bibinfo{pages}{1093} (\bibinfo{year}{2025}).

\bibitem{gedik2025graphene}
\bibinfo{author}{Choi, D.} \emph{et~al.}
\newblock \bibinfo{title}{{Observation of Floquet--Bloch states in monolayer
  graphene}}.
\newblock \emph{\bibinfo{journal}{Nat. Phys.}} \textbf{\bibinfo{volume}{21}},
  \bibinfo{pages}{1100} (\bibinfo{year}{2025}).

\bibitem{GedikOpticalStark2015}
\bibinfo{author}{Sie, E.~J.} \emph{et~al.}
\newblock \bibinfo{title}{{Valley-selective optical Stark effect in monolayer
  {WS}$_2$}}.
\newblock \emph{\bibinfo{journal}{Nat. Mater.}} \textbf{\bibinfo{volume}{14}},
  \bibinfo{pages}{290} (\bibinfo{year}{2015}).

\bibitem{CavalleriGrapheneFloquet2020}
\bibinfo{author}{McIver, J.~W.} \emph{et~al.}
\newblock \bibinfo{title}{{Light-induced anomalous Hall effect in graphene}}.
\newblock \emph{\bibinfo{journal}{Nat. Phys.}} \textbf{\bibinfo{volume}{16}},
  \bibinfo{pages}{38} (\bibinfo{year}{2020}).

\bibitem{Hsieh2021nat}
\bibinfo{author}{Shan, J.-Y.} \emph{et~al.}
\newblock \bibinfo{title}{{Giant modulation of optical nonlinearity by Floquet
  engineering}}.
\newblock \emph{\bibinfo{journal}{Nature}} \textbf{\bibinfo{volume}{600}},
  \bibinfo{pages}{235} (\bibinfo{year}{2021}).

\bibitem{lee2022nat}
\bibinfo{author}{Park, S.} \emph{et~al.}
\newblock \bibinfo{title}{{Steady Floquet--Andreev states in graphene Josephson
  junctions}}.
\newblock \emph{\bibinfo{journal}{Nature}} \textbf{\bibinfo{volume}{603}},
  \bibinfo{pages}{421} (\bibinfo{year}{2022}).

\bibitem{Kogar2024Cr2O3}
\bibinfo{author}{Zhang, X.} \emph{et~al.}
\newblock \bibinfo{title}{{Light-induced electronic polarization in
  antiferromagnetic Cr$_2$O$_3$}}.
\newblock \emph{\bibinfo{journal}{Nat. Mater.}} \textbf{\bibinfo{volume}{23}},
  \bibinfo{pages}{790} (\bibinfo{year}{2024}).

\bibitem{ZhouBPPRL2023}
\bibinfo{author}{Zhou, S.} \emph{et~al.}
\newblock \bibinfo{title}{{Floquet engineering of black phosphorus upon
  below-gap pumping}}.
\newblock \emph{\bibinfo{journal}{Phys. Rev. Lett.}}
  \textbf{\bibinfo{volume}{131}}, \bibinfo{pages}{116401}
  (\bibinfo{year}{2023}).

\bibitem{Waszczak1976PRB}
\bibinfo{author}{Di~Salvo, F.~J.}, \bibinfo{author}{Moncton, D.} \&
  \bibinfo{author}{Waszczak, J.}
\newblock \bibinfo{title}{{Electronic properties and superlattice formation in
  the semimetal TiSe$_{2}$}}.
\newblock \emph{\bibinfo{journal}{Phys. Rev. B}} \textbf{\bibinfo{volume}{14}},
  \bibinfo{pages}{4321} (\bibinfo{year}{1976}).

\bibitem{Robert2006NP}
\bibinfo{author}{Morosan, E.} \emph{et~al.}
\newblock \bibinfo{title}{{Superconductivity in Cu$_{x}$TiSe$_{2}$}}.
\newblock \emph{\bibinfo{journal}{Nat. Phys.}} \textbf{\bibinfo{volume}{2}},
  \bibinfo{pages}{544--550} (\bibinfo{year}{2006}).

\bibitem{jaouen2019phase}
\bibinfo{author}{Jaouen, T.} \emph{et~al.}
\newblock \bibinfo{title}{Phase separation in the vicinity of fermi surface hot
  spots}.
\newblock \emph{\bibinfo{journal}{Phys. Rev. B}}
  \textbf{\bibinfo{volume}{100}}, \bibinfo{pages}{075152}
  (\bibinfo{year}{2019}).

\bibitem{chen2016hidden}
\bibinfo{author}{Chen, P.} \emph{et~al.}
\newblock \bibinfo{title}{{Hidden order and dimensional crossover of the charge
  density waves in TiSe$_2$}}.
\newblock \emph{\bibinfo{journal}{Sci. Rep.}} \textbf{\bibinfo{volume}{6}},
  \bibinfo{pages}{37910} (\bibinfo{year}{2016}).

\bibitem{rohwer2011nature}
\bibinfo{author}{Rohwer, T.} \emph{et~al.}
\newblock \bibinfo{title}{{Collapse of long-range charge order tracked by
  time-resolved photoemission at high momenta}}.
\newblock \emph{\bibinfo{journal}{Nature}} \textbf{\bibinfo{volume}{471}},
  \bibinfo{pages}{490} (\bibinfo{year}{2011}).

\bibitem{li2016nature}
\bibinfo{author}{Li, L.} \emph{et~al.}
\newblock \bibinfo{title}{{Controlling many-body states by the electric-field
  effect in a two-dimensional material}}.
\newblock \emph{\bibinfo{journal}{Nature}} \textbf{\bibinfo{volume}{529}},
  \bibinfo{pages}{185} (\bibinfo{year}{2016}).

\bibitem{kidd2002PRL}
\bibinfo{author}{Kidd, T.}, \bibinfo{author}{Miller, T.},
  \bibinfo{author}{Chou, M.} \& \bibinfo{author}{Chiang, T.-C.}
\newblock \bibinfo{title}{{Electron-hole coupling and the charge density wave
  transition in TiSe$_2$}}.
\newblock \emph{\bibinfo{journal}{Phys. Rev. Lett.}}
  \textbf{\bibinfo{volume}{88}}, \bibinfo{pages}{226402}
  (\bibinfo{year}{2002}).

\bibitem{jaouen2023carrier}
\bibinfo{author}{Jaouen, T.} \emph{et~al.}
\newblock \bibinfo{title}{{Carrier-Density Control of the Quantum-Confined
  1T-TiSe$_{2}$ Charge Density Wave}}.
\newblock \emph{\bibinfo{journal}{Phys. Rev. Lett.}}
  \textbf{\bibinfo{volume}{130}}, \bibinfo{pages}{226401}
  (\bibinfo{year}{2023}).

\bibitem{cercellier2007PRL}
\bibinfo{author}{Cercellier, H.} \emph{et~al.}
\newblock \bibinfo{title}{{Evidence for an excitonic insulator phase in
  1T-TiSe$_2$}}.
\newblock \emph{\bibinfo{journal}{Phys. Rev. Lett.}}
  \textbf{\bibinfo{volume}{99}}, \bibinfo{pages}{146403}
  (\bibinfo{year}{2007}).

\bibitem{monney2011exciton}
\bibinfo{author}{Monney, C.}, \bibinfo{author}{Battaglia, C.},
  \bibinfo{author}{Cercellier, H.}, \bibinfo{author}{Aebi, P.} \&
  \bibinfo{author}{Beck, H.}
\newblock \bibinfo{title}{{Exciton condensation driving the periodic lattice
  distortion of 1T-TiSe$_2$}}.
\newblock \emph{\bibinfo{journal}{Phys. Rev. Lett.}}
  \textbf{\bibinfo{volume}{106}}, \bibinfo{pages}{106404}
  (\bibinfo{year}{2011}).

\bibitem{porer2014NM}
\bibinfo{author}{Porer, M.} \emph{et~al.}
\newblock \bibinfo{title}{{Non-thermal separation of electronic and structural
  orders in a persisting charge density wave}}.
\newblock \emph{\bibinfo{journal}{Nat. Mater.}} \textbf{\bibinfo{volume}{13}},
  \bibinfo{pages}{857} (\bibinfo{year}{2014}).

\bibitem{kim2024origin}
\bibinfo{author}{Kim, K.} \emph{et~al.}
\newblock \bibinfo{title}{{Origin of the chiral charge density wave in
  transition-metal dichalcogenide}}.
\newblock \emph{\bibinfo{journal}{Nat. Phys.}} \textbf{\bibinfo{volume}{20}},
  \bibinfo{pages}{1919} (\bibinfo{year}{2024}).

\bibitem{ishioka2010chiral}
\bibinfo{author}{Ishioka, J.} \emph{et~al.}
\newblock \bibinfo{title}{{Chiral charge-density waves}}.
\newblock \emph{\bibinfo{journal}{Phys. Rev. Lett.}}
  \textbf{\bibinfo{volume}{105}}, \bibinfo{pages}{176401}
  (\bibinfo{year}{2010}).

\bibitem{zhang2021nature}
\bibinfo{author}{Duan, S.} \emph{et~al.}
\newblock \bibinfo{title}{{Optical manipulation of electronic dimensionality in
  a quantum material}}.
\newblock \emph{\bibinfo{journal}{Nature}} \textbf{\bibinfo{volume}{595}},
  \bibinfo{pages}{239} (\bibinfo{year}{2021}).

\bibitem{xu2020nature}
\bibinfo{author}{Xu, S.-Y.} \emph{et~al.}
\newblock \bibinfo{title}{{Spontaneous gyrotropic electronic order in a
  transition-metal dichalcogenide}}.
\newblock \emph{\bibinfo{journal}{Nature}} \textbf{\bibinfo{volume}{578}},
  \bibinfo{pages}{545} (\bibinfo{year}{2020}).

\bibitem{lanzara2024SA}
\bibinfo{author}{Huber, M.} \emph{et~al.}
\newblock \bibinfo{title}{{Ultrafast creation of a light-induced semimetallic
  state in strongly excited 1T-TiSe$_2$}}.
\newblock \emph{\bibinfo{journal}{Sci. Adv.}} \textbf{\bibinfo{volume}{10}},
  \bibinfo{pages}{eadl4481} (\bibinfo{year}{2024}).

\bibitem{Bi2Te3PRB2024}
\bibinfo{author}{Chen, W.} \emph{et~al.}
\newblock \bibinfo{title}{{Distinct light-matter coupling mechanisms in
  Bi$_2$Te$_3$: Crossover from above-gap photoexcitation to light-field
  dressing}}.
\newblock \emph{\bibinfo{journal}{Phys. Rev. B}}
  \textbf{\bibinfo{volume}{110}}, \bibinfo{pages}{L201116}
  (\bibinfo{year}{2024}).

\bibitem{Bi2Te3PRL2025}
\bibinfo{author}{Wang, F.} \emph{et~al.}
\newblock \bibinfo{title}{{Light-field dressing of transient photoexcited
  states above the Fermi energy}}.
\newblock \emph{\bibinfo{journal}{Phys. Rev. Lett.}}
  \textbf{\bibinfo{volume}{134}}, \bibinfo{pages}{146401}
  (\bibinfo{year}{2025}).

\bibitem{mathias2016self}
\bibinfo{author}{Mathias, S.} \emph{et~al.}
\newblock \bibinfo{title}{{Self-amplified photo-induced gap quenching in a
  correlated electron material}}.
\newblock \emph{\bibinfo{journal}{Nat. Commun.}} \textbf{\bibinfo{volume}{7}},
  \bibinfo{pages}{12902} (\bibinfo{year}{2016}).

\bibitem{hedayat2019PRR}
\bibinfo{author}{Hedayat, H.} \emph{et~al.}
\newblock \bibinfo{title}{{Excitonic and lattice contributions to the charge
  density wave in 1T-TiSe$_2$ revealed by a phonon bottleneck}}.
\newblock \emph{\bibinfo{journal}{Phys. Rev. Res.}}
  \textbf{\bibinfo{volume}{1}}, \bibinfo{pages}{023029} (\bibinfo{year}{2019}).

\bibitem{zhou2025CPL}
\bibinfo{author}{Chen, W.} \emph{et~al.}
\newblock \bibinfo{title}{{Revealing Light-Induced Charge Density Wave Melting
  and Carrier Redistribution in 1T-TiSe$_2$}}.
\newblock \emph{\bibinfo{journal}{Chin. Phys. Lett.}}
  \textbf{\bibinfo{volume}{42}}, \bibinfo{pages}{017101}
  (\bibinfo{year}{2025}).

\bibitem{watson2019orbital}
\bibinfo{author}{Watson, M.~D.} \emph{et~al.}
\newblock \bibinfo{title}{{Orbital-and k$_z$-selective hybridization of Se 4p
  and Ti 3d states in the charge density wave phase of TiSe$_{2}$}}.
\newblock \emph{\bibinfo{journal}{Phys. Rev. Lett.}}
  \textbf{\bibinfo{volume}{122}}, \bibinfo{pages}{076404}
  (\bibinfo{year}{2019}).

\bibitem{watson2020strong}
\bibinfo{author}{Watson, M.~D.} \emph{et~al.}
\newblock \bibinfo{title}{{Strong-coupling charge density wave in monolayer
  TiSe$_{2}$}}.
\newblock \emph{\bibinfo{journal}{2D Mater.}} \textbf{\bibinfo{volume}{8}},
  \bibinfo{pages}{015004} (\bibinfo{year}{2020}).

\bibitem{Huang_2021}
\bibinfo{author}{Huang, S.-M.} \emph{et~al.}
\newblock \bibinfo{title}{{Aspects of symmetry and topology in the charge
  density wave phase of 1T-TiSe$_2$}}.
\newblock \emph{\bibinfo{journal}{New J. Phys.}} \textbf{\bibinfo{volume}{23}},
  \bibinfo{pages}{083037} (\bibinfo{year}{2021}).

\bibitem{Zhang2022prb}
\bibinfo{author}{Zhang, R.} \emph{et~al.}
\newblock \bibinfo{title}{{Second-harmonic generation in atomically thin
  1T-TiSe$_2$ and its possible origin from charge density wave transitions}}.
\newblock \emph{\bibinfo{journal}{Phys. Rev. B}}
  \textbf{\bibinfo{volume}{105}}, \bibinfo{pages}{085409}
  (\bibinfo{year}{2022}).

\bibitem{xiao2025arXiv}
\bibinfo{author}{Xiao, T.}, \bibinfo{author}{Huang, T.}, \bibinfo{author}{Bao,
  C.} \& \bibinfo{author}{Sun, Z.}
\newblock \bibinfo{title}{{Interaction Effects on the Electronic Floquet
  Spectra: Excitonic Effects}}.
\newblock \emph{\bibinfo{journal}{arXiv:}} \bibinfo{pages}{2505.07428}
  (\bibinfo{year}{2025}).

\bibitem{Wilson.1978}
\bibinfo{author}{Wilson, J.~A.}, \bibinfo{author}{Barker, A.~S.},
  \bibinfo{author}{Salvo, F. J.~D.} \& \bibinfo{author}{Ditzenberger, J.~A.}
\newblock \bibinfo{title}{{Infrared properties of the semimetal TiSe$_2$}}.
\newblock \emph{\bibinfo{journal}{Phys. Rev. B}} \textbf{\bibinfo{volume}{18}},
  \bibinfo{pages}{2866--2875} (\bibinfo{year}{1978}).

\end{thebibliography}

\begin{addendum}
	\item[Data availability]
	All data supporting the results of this study are available within the article and Supplementary Information. Additional data are available from the corresponding author upon request.
	
  \item [Acknowledgement] 
  This work is supported by the National Natural Science Foundation of China (Grant No.~12234011, 12421004), Tsinghua University Initiative Scientific Research Program (Grant No.~20251080106) National Natural Science Foundation of China (Grant No.~52388201, 92250305, 12327805, 11725418, 11427903), National Key R \& D Program of China (Grant No.~2021YFA1400100), and New Cornerstone Science Foundation through the XPLORER PRIZE.
 
  \item[Author Contributions] 
  S.Z. conceived the research project. F.W. grew the TiSe$_2$ single crystals. F.W. and X.C. performed the TrARPES measurements and analysed the data. T.X. and Z.S. performed the theoretical analysis and calculation. F. W., X. C., C. B. Haoyuan Z., W. C., T. L., T. S., X. T. and Hongyun Z. discussed the results. F.W., X.C., T.X., Z.S. and S.Z. wrote the manuscript, and all authors contributed to the commented on the manuscript.

  \item[Competing Interests] 
  The authors declare that they have no competing financial interests.

\end{addendum}

\begin{figure*}[htbp]
	\centering
	\includegraphics[width=16.8 cm]{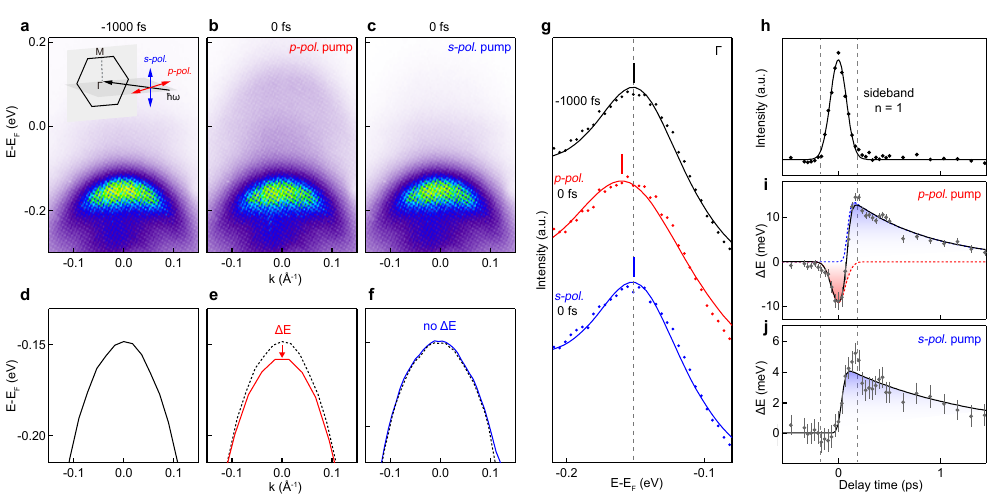}
	\caption*{\textbf{Extended Data Fig.~1 ${\mid}$ Pump polarization dependence of band renormalization.} 
  \textbf{a-c}, Dispersion images measured before pump (\textbf{a}), by \textit{p-pol.} pump (\textbf{b}) and \textit{s-pol.} pump (\textbf{c}) at $\Delta t$ = 0 fs. The pump photon energy is 248 meV and the pump fluence is 2 mJ/cm$^{2}$. The inset shows a sketch of the experimental geometry for the TrARPES setup.
  \textbf{d-f}, Dispersions extracted from \textbf{a-c} by fitting EDCs. 
  \textbf{g}, EDCs at $\Gamma$ point $k$ = 0 for data shown in \textbf{a-c}. 
  \textbf{h}, Intensity of light-field insided sideband as a function of delay time. 
  \textbf{i, j}, Energy shift at $\Gamma$ point as a function of time by \textit{p-pol.} pump (\textbf{i}) and \textit{s-pol.} pump (\textbf{j}). 
  }
	\label{FigS1}
\end{figure*} 

\begin{figure*}[htbp]
	\centering
	\includegraphics[width=16.8 cm]{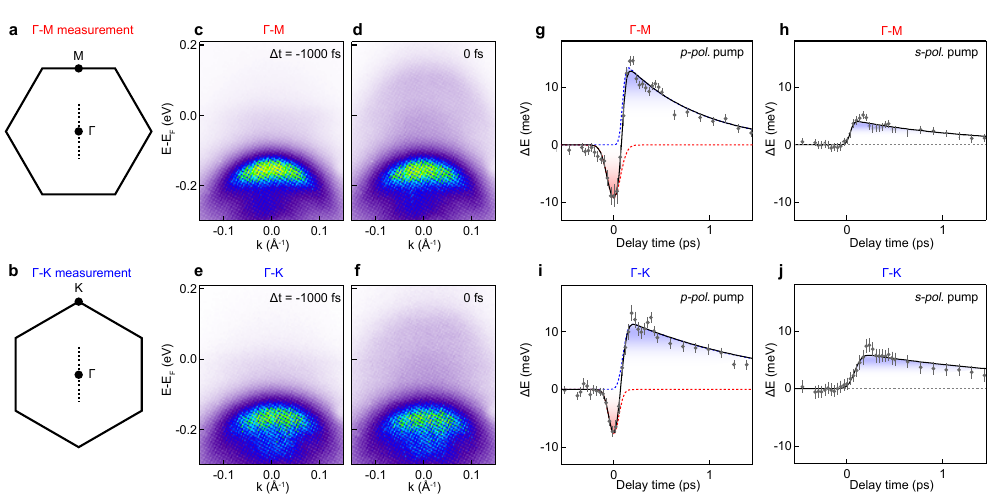}
	\caption*{\textbf{Extended Data Fig.~2 ${\mid}$ Observation of band renormalization along $\Gamma$-M and $\Gamma$-K direction.} 
  \textbf{a,b}, Schematic illustration for the measurement direction (indicated by dashed line).
  \textbf{c-f}, TrARPES dispersion images measured at -1000 fs and 0 fs upon \textit{p-pol.} pump along the $\Gamma$-M direction \textbf{(c,d)} and $\Gamma$-K direction \textbf{(e,f)}. \textbf{g-j}, Energy shift at $\Gamma$ point as a function of delay time by \textit{p-pol.} and \textit{s-pol.} pump measured along the $\Gamma$-M direction \textbf{(g,h)} and $\Gamma$-K direction \textbf{(i,j)}. All data shown in this figure were measured with pump photon energy of 248 meV and pump fluence of 2 mJ/cm$^{2}$.
  }
	\label{FigS2}
\end{figure*}

\end{document}